\documentclass[conference]{IEEEtran}
\IEEEoverridecommandlockouts
\usepackage{cite}
\usepackage{amsmath,amssymb,amsfonts}
\usepackage{algorithmic}
\usepackage{graphicx}
\usepackage{textcomp}
\usepackage{xcolor}
\usepackage{url}
\usepackage{breakurl}
\usepackage{hyperref}
\usepackage{todonotes}
\usepackage{xurl}
\usepackage{amsmath,amssymb,amsfonts}
\usepackage{algorithmic}
\usepackage{graphicx}
\usepackage{multirow}
\usepackage{textcomp}
\usepackage{xcolor}
\usepackage[most]{tcolorbox}
\usepackage{enumitem}
\usepackage{comment}
\usepackage{svg}
\usepackage{pdfpages}

\usepackage{tabularx,booktabs}

\def\BibTeX{{\rm B\kern-.05em{\sc i\kern-.025em b}\kern-.08em
    T\kern-.1667em\lower.7ex\hbox{E}\kern-.125emX}}

\newcommand{\todoblock}[1]{
  \underline{\textbf{TODO}}
}

\newtcolorbox{boxC}[2][]{aibox,title=#2,#1}
\tcbset{
  aibox/.style={
    top=10pt,
    boxrule=0.5pt,
    rounded corners,
    coltitle=black,
    colframe=gray,
    colbacktitle=gray,
    enhanced,
    center,
    attach boxed title to top left={yshift=-0.1in,xshift=0.15in},
    boxed title style={boxrule=0.5pt,colback=black!5!white},
  }
}

\begin{document}

\title{Towards a Small Language Model \\Lifecycle Framework}

\author{\IEEEauthorblockN {Parsa Miraghaei$^1$, Sergio Moreschini$^{1,2}$, Antti Kolehmainen$^1$, and David Hästbacka$^1$}

\IEEEauthorblockA{\textit{$^1$Tampere University}, Tampere, Finland ~ \textit{$^2$University of Oulu}, Oulu, Finland}

\IEEEauthorblockA{parsa.miraghaei@tuni.fi, 
sergio.moreschini@tuni.fi,
antti.kolehmainen@tuni.fi,
david.hastbacka@tuni.fi}
}

\maketitle

\begin{abstract}
\textbf{Background:} The growing demand for efficient and deployable language models has led to increased interest in Small Language Models (SLMs). However, existing research remains fragmented, lacking a unified lifecycle perspective.

\textbf{Objective:} This study aims to define a comprehensive lifecycle framework for SLMs by synthesizing insights from academic literature and practitioner sources.

\textbf{Method:} We conducted a comprehensive survey of 36 works, analyzing and categorizing lifecycle-relevant techniques.

\textbf{Results:} We propose a modular lifecycle model structured into main, optional, and cross-cutting components. The model captures key interconnections across stages, supporting method reuse, co-adaptation, and lifecycle-awareness.

\textbf{Conclusion:} Our framework provides a coherent foundation for developing and maintaining SLMs, bridging theory and practice, and guiding future research and tool development.
\end{abstract}


\section{Introduction}
\label{sec:Intro}

Recent advancements in language models have led to growing interest in Small Language Models (SLMs): compact language models designed for efficiency, deployability, and scalability in resource-constrained environments. Unlike their large-scale counterparts, SLMs are optimized to deliver strong performance while reducing computational and memory overhead. As a result, research has increasingly focused on enabling techniques such as architecture optimization, model compression, knowledge distillation, and deployment.

Despite this momentum, the field remains fragmented. Most of the contributions focus on specific challenges, such as pruning or quantization. However, such challenges are typically developed in isolation, without being incorporated into a comprehensive lifecycle approach. This lack of cohesion complicates both practical implementation and academic positioning. Practitioners must navigate a growing but disjointed ecosystem of tools and techniques, while researchers struggle to situate new work within a structured, repeatable development process.

To address this gap, we propose a comprehensive lifecycle framework for SLMs, grounded in a structured synthesis of academic surveys and practitioner resources. Our goal is to define the fundamental components and interfaces that constitute SLM development, and to clarify how these interact throughout the model lifecycle. The resulting framework categorizes lifecycle stages into main, optional, and cross-cutting components, and highlights their interdependencies to support modularity, adaptability, and reuse.

This work aims to provide both a conceptual foundation and a practical reference for researchers and engineers to develop, deploy, or extend SLMs in a structured and scalable manner.

Section 2 introduces the background. Section 3 presents the methodology. Sections 4 and 5 present the results for RQ1 and RQ2, respectively. Section 6 discusses the results, and Section 7 concludes the work.
\section{Background}
\label{sec:Back}

Recent advancements, including the introduction of models such as Llama-3.2 (1B, 3B, 7B) ~\cite{grattafiori2024llama}, Phi-4 ~\cite{abdin2024phi}, Deepseek-R1 (1.5B, 7B, 8B) ~\ref{guo2025deepseek}, and Qwen2.5 (0.5B, 1.5B, 3B, 7B) ~\cite{yang2024qwen2}, highlight the growing demand for compact yet powerful models that enable low-latency, private, and personalized AI experiences. The rising adoption of SLMs is evident from their increasing download rates on platforms like Hugging Face and the growing number of SLM-focused research initiatives.

Despite their increasing prevalence, definitions of SLMs remains inconsistent. Some frameworks define SLMs as models with fewer than one billion parameters, while others adopt a definition based on task complexity and resource constraints. Recognizing this variability,~\ref{wang2024comprehensive} propose a generalized definition: \textit{"Given specific tasks and resource constraints, we define SLMs as falling within a range where the lower bound is the minimum size at which the model exhibits emergent abilities for a specialized task, and the upper bound is the largest size manageable within limited resource conditions."}


Small Language Models (SLMs) combine computational efficiency with strong task performance, making them ideal for mobile and embedded AI. As shown by Hoffmann et al.\cite{hoffmann2022training}, scaling laws highlight that data quality and volume can outweigh sheer model size, allowing well-trained SLMs to rival much larger models. Beyond standalone use, SLMs play critical roles across the LLM lifecycle—including data curation, alignment, speculative decoding, evaluation, domain adaptation, retrieval-augmented generation, and reasoning. These contributions position SLMs as indispensable components for improving LLM efficiency and scalability~\cite{chen2024role}.

\section{Methodology}
\label{sec:Method}

Guidelines of Garousi et al.~\cite{MLRguidelines}, aiming to integrate insights from both academic and grey literature. However, during the early stages of data collection, we observed a rapid expansion in the volume and relevance of grey sources that were directly aligned with our research focus.

To accommodate this broader and evolving body of evidence, we adapted our approach into a \textbf{comprehensive survey}, allowing for greater flexibility in source inclusion while maintaining structured analysis. This shift enabled a more exhaustive and practice-aware synthesis, enhancing the relevance and coverage of our findings.

Despite this adjustment, our goal remained consistent: to develop a comprehensive, lifecycle-oriented framework for SLMs by synthesizing scholarly and practitioner knowledge. The study is guided by two research questions:
\begin{itemize}
    \item \textbf{RQ1:} How can a comprehensive, end-to-end lifecycle for Small Language Models be defined?
    \item \textbf{RQ2:} What are the key components and interfaces of the SLM lifecycle, and how do they interact?
\end{itemize}
RQ1 focuses on lifecycle structuring, aiming to synthesize diverse contributions into a cohesive framework. RQ2 addresses the dynamic interconnections between components, particularly  modularity, interoperability, and method reuse.

To address these questions, we analyzed 14 core survey papers alongside more than twenty practitioner sources. We extracted lifecycle-related concepts, grouped them into main, optional, and cross-cutting components, and examined their interconnections. This led to a modular and extensible framework that integrates theoretical insights with practical considerations.

We acknowledge that our study diverges from a classic MLR approach, as many of the most relevant sources were identified beyond formal search procedures. These sources, while highly influential in shaping the current understanding and practice of SLM development, are inherently less traceable, limiting the reproducibility of our retrieval process. As a result, we opted not to construct a formal replication package. Instead, we provide transparency through two appendices\footnote{\url{10.6084/m9.figshare.29145146} \label{Package}}: Appendix~A lists all primary sources included in our synthesis, while Appendix~B outlines the full methodological process, including our rationale for shifting toward a comprehensive survey and a reflection on the implications for replicability.
\section{RQ1: How can a comprehensive, end-to-end lifecycle for SLMs be defined?}
\label{sec: RQ1}

\begin{table*}[]
\caption{Coverage of key components in the surveyed literature. Legend: 0 — not covered; m — partially mentioned; 1 — fully covered.}
\label{tab:survey-table}
\resizebox{\linewidth}{!}{%
\begin{tabular}{lll|l|l|l|l|l|l|l|l|l|l|l|l|l}
\hline
Paper &  &
\ref{wang2024comprehensive} &
\ref{van2024survey} &
\ref{zhu2024survey} &
\ref{tang2024survey} &
\ref{bai2024beyond} & 
\ref{wan2023efficient} & 
\ref{wang2025empowering} & 
\ref{johnson2025improving} & 
\ref{yuan2024llm} &
\ref{xu2024device} &
\ref{subramanian2025small} & 
\ref{lu2024small} & 
\ref{miao2023towards} & 
\ref{wang2024model} \\ \hline
                      & Efficient Architectures & 1 & 1 & 0 & 1 & 1 & 1 & m & 1 & 1 & 1 & m & 1 & 1 & 1 \\ 
                      & Merging & 0 & 1 & 0 & 0 & 1 & m & 0 & 0 & 0 & m & m & 0 & 0 & m \\ 
Initiation            & Factorization & 1 & 0 & 1 & 1 & 1 & 1 & 1 & m & 1 & 1 & 0 & 0 & 0 & 1 \\ 
                      & Pruning & 1 & 1 & 1 & 1 & 1 & 1 & 1 & 1 & 1 & 1 & m & 0 & 1 & 1 \\ 
                      & NAS & 0 & 0 & m & 1 & 1 & 0 & 0 & 1 & m & 0 & 0 & m & 0 & 1 \\ 
                      & Continual Pretraining & 1 & 0 & 0 & 0 & 0 & 0 & 0 & 0 & 0 & 0 & m & 0 & 0 & 0 \\ 
                      & Pruning and Distillation & 0 & 1 & 0 & 0 & 0 & 0 & 0 & 0 & 0 & 0 & 0 & 0 & 0 & 0 \\ \hline
Distillation          &  & 1 & 1 & 1 & 1 & 1 & 1 & 1 & 1 & 0 & 1 & 1 & m & 1 & 1 \\ \hline
General Lifecycle     & Transformation I & 1 & 0 & 0 & 0 & m & m & m & 0 & 0 & 0 & 0 & 0 & 0 & 0 \\
                      & Transformation II & 1 & 0 & 0 & 0 & 0 & 1 & m & 0 & m & 0 & 0 & m & 0 & 0 \\ \hline
Quantization          &  & 1 & 1 & 1 & 1 & 1 & 1 & 1 & 1 & 1 & 1 & 1 & m & 1 & 1 \\ \hline
Deployment            &  & 1 & 0 & 0 & 0 & 1 & 1 & 1 & 1 & 1 & 1 & 0 & 1 & 1 & 1 \\ \hline

On-Device Learning      &  & 1 & 0 & 0 & 0 & 0 & 0 & 0 & 0 & m & 1 & m & m & 0 & 0 \\ \hline
Federated learning      &  & 0 & 0 & 0 & 0 & 1 & 0 & m & 0 & 0 & 0 & 0 & 0 & 0 & 0 \\ \hline
                        
Efficient Tuning        &  & 1 & 0 & 1 & 0 & 1 & 1 & 0 & 1 & 0 & 0 & 1 & m & 1 & 1 \\ \hline
Data Selection          &  & 1 & 1 & m & m & 1 & 1 & 0 & 0 & 0 & 0 & 1 & m & 0 & 1 \\ \hline
Evaluation              &  & 1 & 1 & 1 & 0 & 1 & 0 & 0 & 0 & 1 & 0 & 0 & 1 & 0 & 0 \\ \hline

\end{tabular}%
}
\end{table*}

Recent studies on SLMs have made significant progress in addressing individual stages of model development, including efficient architecture design, quantization, pruning, distillation, and deployment. However, the literature remains still largely fragmented. In Table~\ref{tab:survey-table} we provide an overview of studies that specifically target small, efficient, or compressed language models and address at least five core components of the SLM lifecycle.

\ref{wang2024comprehensive} offer the most extensive treatment to date, covering a wide range of techniques relevant to SLMs, including initialization strategies, quantization methods, distillation, and efficient deployment. While they discuss lifecycle stages in general terms, their work primarily catalogues individual techniques and stops short of articulating the interconnections or dependencies between these stages. Similarly, \ref{wan2023efficient} and \ref{tay2022efficient} provide deep dives into efficiency techniques—such as pruning, knowledge distillation, and architecture design—but do not contextualize them within a structured, reusable lifecycle model.

Other surveys provide strong technical overviews of individual lifecycle components. For instance, \ref{bai2024beyond} and \ref{yuan2024llm} examine inference optimization from both algorithmic and hardware perspectives, while \ref{van2024survey} focus on initialization, compression, and deployment in edge environments. However, these contributions generally lack a holistic framework to guide how such techniques are integrated throughout the model development process. \ref{xu2024device} further specialize in on-device learning, proposing practical constraints and hardware-aware trade-offs, yet they treat this as a standalone capability rather than a lifecycle-integrated module.

Distillation has been extensively studied across surveys such as \ref{johnson2025improving} and \ref{xu2024device}, with a focus on taxonomies of white-box and black-box approaches.
Similarly, federated learning is addressed by \ref{bai2024beyond} and \ref{wang2025empowering} as a privacy-preserving training paradigm, yet its role within a broader SLM lifecycle is left unexplored.

A few works, such as \ref{wang2024comprehensive}, \ref{wan2023efficient}, touch on lifecycle transformations in LLMs, particularly highlighting Transformation I and Transformation II processes. However, these discussions are abstract and not specialized for the unique constraints and opportunities presented by SLMs. Furthermore, while \ref{wang2024comprehensive} ,\ref{bai2024beyond} explore Parameter-Efficient Fine-Tuning (PEFT) in depth, they largely omit how PEFT interacts with quantization or distillation, or how it might influence component-aware deployment strategies.

Evaluation practices are extensively discussed in works such as~\ref{bai2024beyond}, \ref{wang2024comprehensive}, \ref{cheng2024small}, particularly in relation to benchmarks, trustworthiness, and evaluation metrics. However, these studies often address these aspects in isolation and typically overlook key issues such as benchmark limitations and the role of automated evaluation methods. Similarly, \ref{wan2023efficient}, \ref{bai2024beyond} provide a comprehensive taxonomy of data selection methods, but do not clarify how such techniques are reused across stages such as distillation.

Building on the gaps identified in the current literature, we define a comprehensive lifecycle for SLMs as a modular, extensible framework composed of main, optional, and cross-cutting components, all of which are interconnected. While existing surveys provide deep insights into individual techniques, they often treat these in isolation, lacking a unifying perspective on how they interact across the model development and deployment pipeline.

Our proposed lifecycle addresses this gap by structurally organizing the development process into:
\begin{itemize}
    \item \textbf{5 Main components:} Initialization, Distillation, General lifecycle, Quantization, and Deployment - which reflect the core stages of SLM development
    \item \textbf{2 Optional component:} On-device Learning, and Federated Learning - increasingly relevant for privacy-preserving and resource-constrained environments
    \item \textbf{4 Cross-cutting components:} Data Selection, Evaluation, Efficient Fine Tuning, and Inference Optimization - which influence and support multiple stages of the lifecycle
\end{itemize}

Crucially, this lifecycle is not linear. The components are \textbf{interconnected}, allowing methodologies and insights developed in one phase (e.g., quantization-aware fine-tuning, PEFT-aware pruning) to influence or adapt to the constraints and objectives of others. This interconnectedness ensures that the lifecycle remains flexible, reusable, and responsive to both architectural innovations and practical deployment considerations. Thus, the lifecycle we propose moves beyond a collection of best practices and toward a coherent, system-level view of SLM development—capable of guiding both research and implementation in a structured and scalable manner.

\section{RQ2: What are the key components and interfaces of the SLM lifecycle, and how do they interact?}
\label{sec:RQ2}

\begin{figure*}[t!]
    \centering
    \includegraphics[width=.87\textwidth]{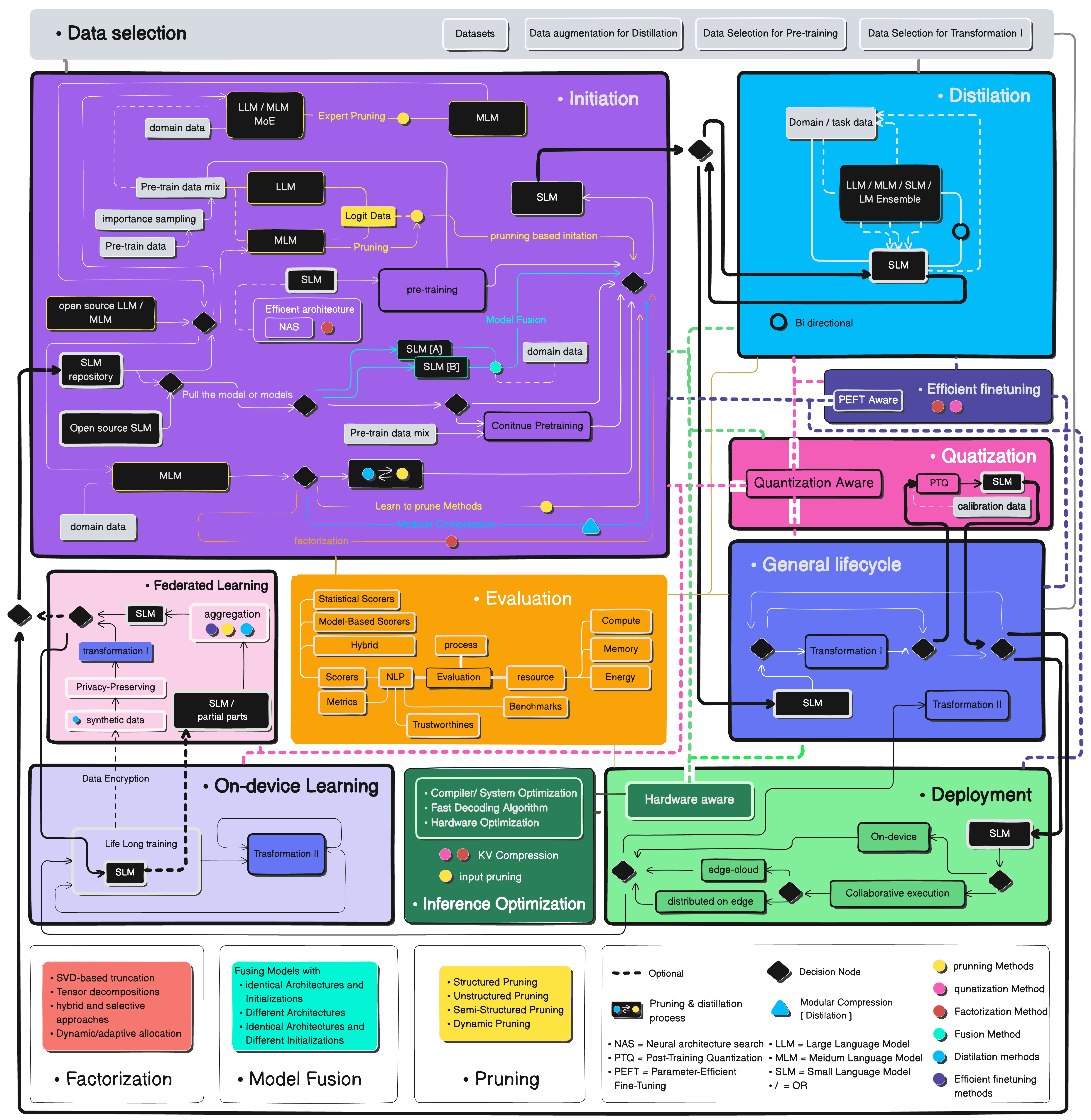}
    \caption{SLM lifecycle}
    \label{fig:slm-lifecycle}
    \vspace{-5mm}
\end{figure*}

Figure~\ref{fig:slm-lifecycle} illustrates our proposed lifecycle for SLMs, This schematic captures the entire process comprehensively in one integrated structure. It emphasizes the iterative nature of model development, showcases relationships among components, differentiates clearly between main, optional, and cross-cutting components, and clarifies significant interdependencies across components within the lifecycle.

\subsection{Main Components}
\subsubsection{Initialization} The initialization phase involves selecting or developing a suitable SLM based on specific goals and evaluation criteria. As illustrated in Figure \ref{fig:slm-lifecycle}, this process can follow three main pathways: (1) designing an efficient model from scratch, (2) compressing a larger model, or (3) initializing from existing models available internally or through open-source repositories, with potential for further pretraining or merging.

\begin{itemize}
    \item \textbf{Efficient Model Design:} Efficiency-oriented SLMs aim to reduce compute, memory, and energy costs without compromising performance. Recent innovations include attention optimizations such as Multi-Query Attention (MQA), Grouped Query Attention (GQA), and Multi-Head Latent Attention (MLA), which reduce key-value cache usage. Parameter sharing (e.g., shared FFNs in MobiLlama, repeated blocks in MobileLLMs) and embedding reuse further enhance compactness. Lightweight alternatives to standard components—such as SiLU/SwiGLU activations and RMSNorm normalization—offer both performance and resource gains~\ref{wang2024comprehensive}. Efficient self-attention mechanisms (e.g., kernelization, fixed or learnable sparsity patterns) and long-context strategies (e.g., positional extrapolation, sliding windows, memory-augmented attention) extend capabilities without quadratic scaling~\ref{van2024survey},~\ref{wan2023efficient},~\ref{wang2024comprehensive}. Sparse Models and Conditional Computation models sparsely activate a subset of the parameters (e.g MoE-based LLMs), improving the parameter to FLOPs ratio MoE-based LLMs~\ref{wan2023efficient}. In addition,~\ref{tay2022efficient} highlights complementary techniques such as Combination of Patterns (CP), learnable side memory modules, and downsampling strategies for improving model efficiency. Emerging architectures such as State Space Models (e.g., Mamba) and sequential models (e.g., RWKV, Hyena) provide efficient alternatives to transformers~\ref{wang2024comprehensive},~\ref{van2024survey}. Automated design techniques like Neural Architecture Search (NAS) further streamline model construction by optimizing for task-specific or hardware-specific constraints. 
    
    \item \textbf{Model Initialization Techniques Using Larger Models:} There are several ways to derive smaller models from larger ones. Common strategies include: (1) Pruning: Pruning techniques vary by sparsity pattern, timing, and hardware impact~\ref{cheng2024survey}. Unstructured pruning removes individual weights for fine-grained sparsity, while semi-structured pruning introduces patterns (e.g., stripes or blocks) to improve hardware efficiency. Structured pruning eliminates entire components like filters or attention heads, enabling universal speedups on standard hardware~\ref{zhu2024survey}. Dynamic pruning adapts the network at inference based on input, generating different subnetworks per instance, including NAS-based methods~\ref{johnson2025improving},~\ref{bai2024beyond},~\ref{xu2024device},~\ref{van2024survey}. Pruning can also be static, applied before, during, or after training to yield a fixed subnetwork. In sparse MoE models, pruning targets less active experts. Learn-to-prune approaches train models or meta-learners to identify redundant parts using sparsity regularization, meta-learning, GNNs, or RL. (2) Iterative pruning and distillation combine pruning with repeated knowledge distillation to compress models while preserving performance.(3) Modular Compression is a distillation-based method that compresses transformer models by replacing modules with compact substitutes during training. It enables efficient learning without extra loss terms~\ref{tang2024survey},~\ref{xu2020bert}.
    (4) Low-Rank Factorization compresses models by decomposing weight matrices to reduce parameters and computation. While prior surveys lack clear categorization, we group methods into four families, including SVD-based, tensor decompositions, hybrid/selective, and adaptive rank allocation~\ref{johnson2025improving},~\ref{wang2024comprehensive}.
    
    \item \textbf{Model Fusion:} involves training separate SLMs for different capabilities and then merging them—a concept highlighted in~\ref{subramanian2025small} and exemplified by MergeKit~\ref{goddard2024mergekit}. MergeKit is a library designed to combine language model checkpoints for improved performance and flexibility. It supports methods like linear interpolation, SLERP, TIES, and DARE for models with shared architectures and initializations, and techniques like Git-Rebasin and Optimal Transport Fusion for those with shared architectures but different initializations. It also explores cross-architecture fusion with CALM and FUSELLM, and introduces novel approaches like FrankenMerging and Franken-MoE to construct new model architectures beyond direct parameter merging.

    \item \textbf{Continual Pretraining:} extends initial training on domain-specific data, enabling SLMs to gain specialized knowledge and excel in fields like healthcare, science, finance, and law~\ref{wang2024comprehensive}.
    
\end{itemize}

\begin{figure*}[t!]
    \centering
    \includegraphics[width=0.8\textwidth]{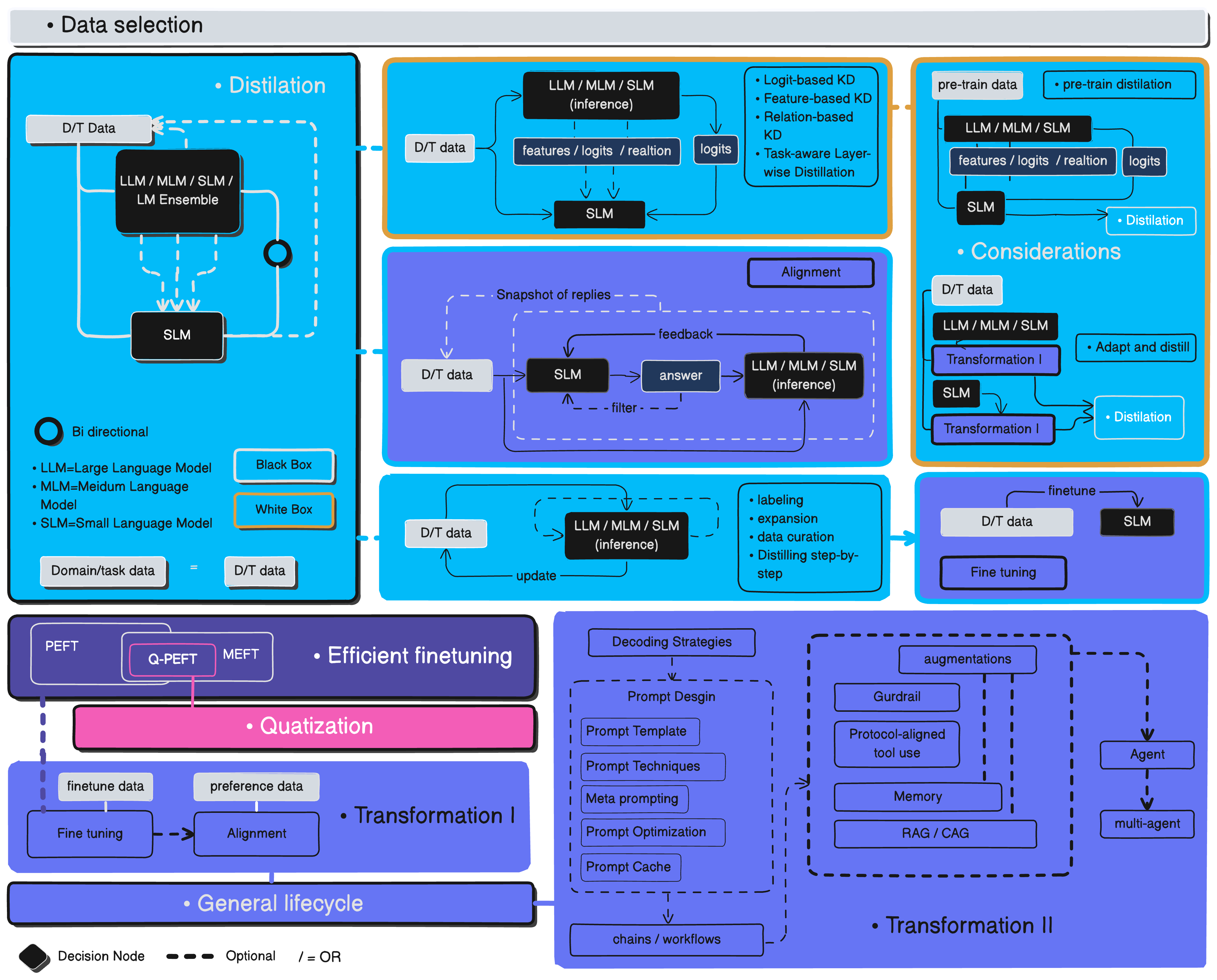}
    
    \caption{This figure provides a more detailed view of the Distillation, Efficient Finetuning, General lifecycle components that are only briefly mentioned in Figure \ref{fig:slm-lifecycle} }
    \label{fig:zoomin}
    \vspace{-5mm}
\end{figure*}

\subsubsection{Distillation} Knowledge distillation transfers learned representations from a large, often resource-intensive teacher model to a smaller, more efficient student model. Figure~\ref{fig:zoomin} provides a high-level overview of this component. Distillation strategies generally fall into two broad categories:
\begin{itemize}
    \item \textbf{White-box distillation} leverages full access to a teacher model’s internal states, enabling the student to learn from richer signals beyond final outputs. Logit-based distillation matches softened teacher outputs to convey inter-class relationships, while feature-based distillation transfers intermediate representations for deeper structural guidance. Response-based distillation focuses only on final predictions and is typically less effective, whereas relation-based distillation aligns structural relationships across layers or samples~\ref{johnson2025improving}. The choice of architecture also matters: same-architecture distillation uses scaled-down versions of the teacher, cross-architecture allows flexible student designs, and ensemble distillation aggregates outputs from multiple teachers to enhance robustness and generalization~\ref{johnson2025improving}. White-box methods address domain gaps between teacher and student using techniques like Adapt-and-Distill, Domain Adaptation, and Pre-train Distillation to align training distributions or tasks, especially when using domain-specific data or task conditioning~\ref{wang2024comprehensive}.

    \item \textbf{Black-box distillation} transfer knowledge from a teacher model without accessing its internals, relying instead on observable input-output behavior \ref{xu2024survey},~\ref{yuan2024llm},~\ref{wang2025empowering}. Students may learn from direct teacher outputs (labeling), teacher-generated prompts (expansion), synthetic data based on metadata (curation), or teacher feedback (correction and ranking). In some cases, students generate outputs themselves and filter them for quality (self-knowledge). These methods are typically combined with Supervised Fine-Tuning for initial imitation learning, and advanced techniques use Reinforcement Learning or Rank Optimization to align student outputs with teacher preferences, enabling efficient adaptation without compromising effectiveness.
\end{itemize}

\subsubsection{General lifecycle} The general lifecycle, though foundational to both SLMs and LLMs, has been underexplored in existing surveys. While~\ref{wang2024comprehensive} recognize its importance, their emphasis diverges from ours. Drawing on~\ref{zhao2023survey} and contextualized by~\ref{bai2024beyond} and~\ref{cheng2024small}, we propose a structured abstraction based on two core transformations, as outlined in Figure~\ref{fig:zoomin}.
\begin{itemize}
    \item \textbf{Transformation I} refers to weight-level modifications that alter a model’s parameters. This includes task- and domain-specific fine-tuning methods like Supervised Fine-Tuning (SFT), Reinforcement Fine-Tuning (RFT)~\ref{guo2025deepseek}, instruction tuning, learning-to-reason, and in-context learning , as well as alignment techniques such as Reinforcement Learning from Human Feedback (RLHF), Reinforcement Learning from AI Feedback (RLAIF), using Proximal Policy Optimization (PPO), Direct Preference Optimization (DPO), and Kahneman–Tversky Optimization (KTO), which use preference-based feedback to align models with human expectations~\ref{zhao2023survey},~\ref{hadi2023large}.
    \item \textbf{Transformation II} involves inference-time techniques that improve model performance without changing weights. Despite their simplicity, these methods can significantly enhance outcomes~\ref{lu2024small}. This transformation draws from recent advancements synthesized in~\ref{minaee2025largelanguagemodelssurvey},\ref{Dong2024inference} and includes four main categories. (1) Independent Self-Improvement focuses on enhancing performance using the model’s frozen weights through decoding strategies such as greedy search, beam search, and top-k/top-p sampling, as well as techniques like sampling multiple candidate generations and isolating influential layers or neurons. (2) Context-Aware Self-Improvement involves prompt engineering methods like Chain of Thought, Tree of Thought, Self-Consistency, and Reflection, along with augmentation techniques such as Retrieval-Augmented Generation (RAG) and knowledge graph integration to enhance contextual relevance. (3) Model-Aided Self-Improvement introduces chaining, tool use, and support from smaller or specialized models (e.g., Amateur LMs, Reward Models) to guide reasoning and output generation. (4) agent-based systems such as ReAct, ReWOO, and collaborative frameworks like AutoGen enable modular, iterative, and interactive decision-making during inference. Additional advances like Prompt Optimization, Continuous Prompt Optimization, and Knowledge Graph Augmentation are also highlighted in\ref{zhao2023survey}.
\end{itemize}

\subsubsection{Quantization} Quantization reduces storage and computation in language models by converting floating-point values to integers or discrete forms, with minimal impact on accuracy—making it essential for SLM deployment. It includes Quantization-Aware Training (QAT) and Post-Training Quantization (PTQ); while PTQ is emphasized here, QAT is considered an interconnected layer and will be discussed in the corresponding section.

According to the taxonomy in~\ref{yuan2024llm}, PTQ techniques are classified into three types: weight-only quantization (e.g., AWQ, GPTQ), very low-bit quantization (e.g., BiLLM, PB-LLM), and joint weight-activation quantization (e.g., OliVe, SmoothQuant), each balancing compression and performance differently.

\subsubsection{Deployment} refers to the strategies used to run language models efficiently across different environments. For Small Language Models (SLMs), this typically involves either on-device inference, which runs directly on a single device (e.g., smartphones or IoT devices) for low-latency and offline use, or collaborative execution, which distributes inference across multiple units—either through distributed deployment or cloud-edge collaboration—to optimize scalability, responsiveness, and resource use~\ref{wang2025empowering},\ref{wang2024comprehensive},\ref{xu2024device}. Effective deployment, especially on-device, depends heavily on selecting suitable hardware accelerators—such as CPUs, GPUs, or NPUs—that balance performance, energy efficiency, and cost, as these choices directly influence model responsiveness and efficiency~\ref{xu2024device}.

\subsection{Optional Components}
\subsubsection{On-device Learning} enables small language models (SLMs) to adapt using locally available data without external transmission, ensuring privacy. While prior work emphasizes retrieval-augmented generation and on-device fine-tuning as core paradigms ~\ref{wang2024comprehensive}, we advocate for a more holistic view that embeds local updates within the full lifecycle of SLMs. This integration supports mechanisms such as local alignment, prompt updates, and continual adaptation, aligning with continual learning principles outlined in~\ref{biesialska2020continual}.

Rather than viewing on-device learning as an isolated capability, we argue that local knowledge updates should be embedded within the general lifecycle framework introduced in earlier sections. This perspective enables the incorporation of additional processes such as local alignment, prompt and chain updates, and continual adaptation over time. These mechanisms align closely with the principles of continual learning, as emphasized in~\ref{biesialska2020continual}, and underscore the importance of treating on-device learning as a dynamic, lifelong process. To address the realities of resource-constrained settings, we propose an environment-aware SLM lifecycle that balances efficiency with ongoing local learning. Practical strategies identified in~\ref{xu2024device}—including quantization-aware scaling, sparse updates, and lightweight training engines—highlight how adaptation can be achieved with minimal overhead. 

\subsubsection{Federated Learning}
Federated Learning (FL) enables decentralized training across multiple devices without centralizing sensitive data, making it particularly well-suited for SLMs operating in edge environments~\ref{bai2024beyond}. Recent advances show that FL is increasingly feasible for LLMs on modern edge infrastructures~\ref{bai2024beyond}, with methods like FedSpaLLM supporting sparse model optimization to reduce communication and computation costs, and SplitLoRA enabling parameter-efficient fine-tuning by distributing model segments across devices. Beyond weight sharing, federated distillation allows devices to exchange output distributions or intermediate representations instead of parameters, further minimizing communication while preserving privacy~\ref{li2024federated}. Complementary to this, data synthesis techniques generate artificial yet semantically meaningful samples to train local models without exposing sensitive user data~\ref{shao2023survey}. Collectively, these approaches adapt FL to the needs of SLMs, enabling scalable, privacy-preserving model training in decentralized settings.

\subsection{Cross-cutting Components}
\subsubsection{Data Selection}
Data selection, or data efficiency~\ref{bai2024beyond}, is a core challenge in machine learning focused on identifying optimal data subsets to maximize objectives such as model performance, computational efficiency, evaluation fidelity, and mitigation of harmful behaviors like bias and toxicity~\ref{wan2023efficient},~\ref{albalak2024survey}. The ideal dataset closely matches the target distribution for evaluation. In the lifecycle of LLMs and SLMs, data selection is tailored to different training stages: efficient pretraining filters large corpora for samples that best support generalization, using methods like SSPT, DSIR, and DoReMi, while downstream selection, including instruction tuning and fine-tuning~\ref{albalak2024survey}, targets high-quality, task-specific data—leveraging techniques such as Instruction Mining, TS-DShapley, and LTD Instruction Tuning, with evidence from LIMA and AlpaGasus that small, curated sets can yield strong results. Distillation strategies like Distilling Step-by-Step further highlight the impact of targeted selection by focusing on reasoning traces rather than simple input-output pairs. Community-curated datasets~\ref{liu2024datasets} embody the cumulative value of prior selection, serving as reusable, high-quality resources for efficient model development. In summary, effective data selection across the entire model lifecycle underpins performance, efficiency, and ethical outcomes.

\subsubsection{Evaluation} Evaluation is a foundational aspect of the language model lifecycle, providing systematic assessment of performance, trustworthiness, and efficiency across diverse tasks and operational contexts. Benchmark suites such as MMLU and HellaSwag measure core capabilities but are vulnerable to data contamination, making careful curation and transparent reporting essential~\ref{lu2024small},~\ref{wang2024comprehensive},~\ref{zhu2024survey},~\ref{yuan2024llm}. Trustworthiness evaluation covers robustness to adversarial inputs, privacy protection, reliability (including hallucination and consistency), and safety concerns such as toxicity and bias~\ref{wang2024comprehensive},~\ref{yuan2024llm}, all of which are vital for user-facing or high-stakes deployments. Resource efficiency—spanning computational cost, memory, energy, and deployment overhead—is particularly important for SLMs and shapes deployment strategies in constrained environments~\ref{bai2024beyond},~\ref{yuan2024llm}. Automated evaluation methods range from statistical scorers like BLEU and ROUGE to model-based and hybrid approaches, with the latter providing stronger alignment with human judgment and greater scalability~\ref{ip2025llm},~\ref{microsoft2024llmeval}. Ultimately, evaluation should be an integrated, continuous process that informs model iteration, balances performance with sustainability and safety, and supports real-world usability at scale.

\subsubsection{Efficient Fine Tuning}
Efficient fine-tuning is central to recent surveys on Small and Large Language Models, with Parameter-Efficient Fine-Tuning (PEFT) methods designed to minimize trainable parameters and memory requirements while retaining performance comparable to full fine-tuning. PEFT strategies are typically classified as additive fine-tuning, which adds task-specific modules; partial fine-tuning, which adapts only select parameters; reparameterized fine-tuning, exemplified by low-rank techniques like LoRA; hybrid fine-tuning, which integrates multiple strategies; and unified fine-tuning: Presents a unified framework for incorporating diverse fine-tuning methods, often using a single PEFT method with mechanisms like masking or sharing across layers~\ref{xu2023parameter}. By lowering resource demands without compromising adaptability, PEFT enables scalable fine-tuning in resource-constrained, rapid deployment, and personalized application scenarios, making it a critical part of the SLM lifecycle.

\subsubsection{Inference Optimization}

Inference optimization is central to scaling both LLMs and SLMs for efficient, real-time use, spanning compiler and system improvements, decoding algorithm advances, and hardware-level innovations~\ref{yuan2024llm}. Compiler and system optimizations, such as operator fusion (e.g., FlashAttention, DeepSpeed-Inference, TensorRT-LLM), and dynamic memory management frameworks like PagedAttention and FlexGen, minimize latency and memory bottlenecks, enabling faster and more scalable inference, even on memory-limited devices; parallel serving systems like ORCA and LightLLM further improve hardware utilization and multi-user latency. Decoding optimizations—including early exiting (CONFIDENT, SkipDecode), contextual sparsity (Deja Vu), mixture-of-experts (Sparse Mixer), and multi-token generation strategies (Skeleton-of-Thoughts, Lookahead Decoding)—reduce computation per token and accelerate generation, though challenges remain for non-autoregressive methods with decoder-only models. At the hardware level, spatial computing (Groq LPU), processing-in-memory (HBM-PIM), and innovations in data formats (FP8 precision, variable-length and bit-sharing encoding, outlier-aware quantization) cut memory costs and improve energy efficiency, making high-performance inference feasible across deployment scenarios.

\subsection{Interconnections within the Lifecycle}

Interconnection within the lifecycle of SLMs hinges on two key principles:

\subsubsection{Component Awareness and Lifecycle Adaptation}
This is exemplified in practices like Quantization-Aware Training (QAT), which ensures that stages such as pretraining, distillation, and parameter-efficient finetuning (PEFT) account for quantization constraints. Mixed-precision methods like OneBit and BitNet b1.58 embed quantization into pretraining, while distillation methods such as LLM-QAT use data-free techniques to train quantized student models from full-precision teachers ~\ref{yuan2024llm}, \ref{wang2024comprehensive}. Beyond quantization, hardware-awareness is critical—SLMs must be tailored to memory, compute, and accelerator constraints across edge, cloud, or distributed setups. This influences data formats (e.g., low-bit floating point, variable-length encoding) and may necessitate specialized architectures like spatial designs seen in Groq’s LPU or Graphcore’s IPU, which enhance inference efficiency by minimizing memory interactions~\ref{yuan2024llm},\ref{wang2019haq}. Another key interconnection is PEFT-awareness—evident in methods like LoRAPrune~\ref{zhang2023loraprune}, and platforms like LoraHub~\ref{xu2023parameter} or device-optimized LoRA libraries~\ref{xu2024device}, which influence pruning, fusion, and deployment. Multi-LoRA~\ref{thomas2024multi}, in particular, supports efficient deployment by coordinating multiple finetuning modules across scenarios. Together, these examples illustrate how awareness across components leads to reciprocal design choices and optimized lifecycle strategies.

\subsubsection{Methodological Transfer Across Components}
lies in the reuse of core techniques across different stages of the language model lifecycle, illustrating a cohesive design philosophy. For instance, factorization—central to LoRA for parameter-efficient fine-tuning—is also leveraged in inference-time KV compression, while quantization plays a dual role in both Q-PEFT and speeding up inference. Distillation, initially intended for model compression, has been effectively adapted to support decentralized training in federated learning~\ref{li2024federated}. Likewise, dynamic pruning extends beyond compression, enabling input pruning in fast decoding pipelines to enhance inference efficiency~\ref{yuan2024llm}.

\section{Discussion}
\label{sec:Disc}

The lifecycle model presented in this paper represents an essential step toward establishing a foundational structure for managing Small Language Models (SLMs). Although not intended as a prescriptive operational guide, it provides the necessary framework upon which modular and flexible operational strategies can be developed. Simple yet strategic operational adjustments—such as tuning batch size, model dimensions, attention mechanisms, or communication protocols—can lead to substantial gains in performance and cost-efficiency, as demonstrated in~\cite{ashkboos2024computational} and further emphasized by~\cite{hu2024minicpm}. By clarifying how these operational enhancements can systematically improve SLM performance, our lifecycle model supports teams in creating adaptive pipelines tailored to evolving needs. Additionally, this lifecycle model serves as a shared reference, allowing researchers to contextualize innovations, identify gaps, and build upon existing contributions. Thus, it not only advances operational efficiency but also enriches the broader research landscape. Ultimately, this paper marks a critical step in defining Small Language Model Operations (SLMOps), bridging traditional ML operations~\cite{Kreuzberger2023MLOps,moreschi2023mlops} and LLM workflows~\cite{llmops_survey_search,sinha2024llmops}, and paving the way toward the growing integration of SLMs within compound AI systems~\cite{compound-ai-blog}.
\section{Conclusion}
\label{sec:Conclusion}

This paper presented a lifecycle framework for Small Language Models (SLMs), addressing the current fragmentation in the literature and practice. By synthesizing insights from 36 sources, we identified and organized the key stages of SLM development into main, optional, and cross-cutting components. The resulting framework clarifies the interconnections between lifecycle stages and highlights opportunities for method reuse and co-adaptation.

Our findings contribute to both research and practice. For researchers, the lifecycle provides a conceptual map that situates individual techniques within a broader, integrated workflow. For practitioners, it offers a practical guide for designing scalable and efficient SLM pipelines, particularly in resource-constrained or modular deployment settings.

The lifecycle also lays the groundwork for future research into specialized operational frameworks, such as SLMOps, and the systematic evaluation of SLM engineering strategies across domains. As compact and efficient language models continue to evolve, having a clear, lifecycle-aware foundation will be essential to support their sustainable development and adoption.

\textbf{Acknowledgment:} This work was supported by the Industry X and 6GSoft projects funded byBusiness Finland.

\section*{Declaration of generative AI and AI-assisted technologies in the writing process}
During the preparation of this work the author used ChatGPT in order to  improve language and readability. After using this service, the authors reviewed and edited the content as needed and take full responsibility for the content of the publication.

\bibliographystyle{elsarticle-num} 
\bibliography{refs2.bib}

\newpage

\section*{Appendix A: Original Studies (OS$_s$)} 
\label{The Original Studies}

{\footnotesize
  \begin{enumerate}[labelindent=-5pt,label={[OS}{\arabic*]}]

\item \label{guo2025deepseek}
Guo, D., Yang, D., Zhang, H., Song, J., Zhang, R., Xu, R., ... \& He, Y. (2025). Deepseek-r1: Incentivizing reasoning capability in llms via reinforcement learning. arXiv preprint arXiv:2501.12948.
  
\item \label{wang2024comprehensive}
Wang, F., Zhang, Z., Zhang, X., Wu, Z., Mo, T., Lu, Q., ... \& Wang, S. (2024). A comprehensive survey of small language models in the era of large language models: Techniques, enhancements, applications, collaboration with llms, and trustworthiness. arXiv preprint arXiv:2411.03350.

\item \label{wan2023efficient}
Wan, Z., Wang, X., Liu, C., Alam, S., Zheng, Y., Liu, J., ... \& Zhang, M. (2023). Efficient large language models: A survey. arXiv preprint arXiv:2312.03863.

\item \label{tay2022efficient}
Tay, Y., Dehghani, M., Bahri, D., \& Metzler, D. (2022). Efficient transformers: A survey. ACM Computing Surveys, 55(6), 1-28.

\item \label{bai2024beyond}
Bai, G., Chai, Z., Ling, C., Wang, S., Lu, J., Zhang, N., ... \& Zhao, L. (2024). Beyond efficiency: A systematic survey of resource-efficient large language models. arXiv preprint arXiv:2401.00625.

\item \label{yuan2024llm}
Yuan, Z., Shang, Y., Zhou, Y., Dong, Z., Zhou, Z., Xue, C., ... \& Keutzer, K. (2024). Llm inference unveiled: Survey and roofline model insights. arXiv preprint arXiv:2402.16363.

\item \label{van2024survey}
Van Nguyen, C., Shen, X., Aponte, R., Xia, Y., Basu, S., Hu, Z., ... \& Nguyen, T. H. (2024). A survey of small language models. arXiv preprint arXiv:2410.20011.

\item \label{xu2024device}
Xu, J., Li, Z., Chen, W., Wang, Q., Gao, X., Cai, Q., \& Ling, Z. (2024). On-device language models: A comprehensive review. arXiv preprint arXiv:2409.00088.

\item \label{johnson2025improving}
Johnson, A., Gonzalez, M., Kim, D., Sharma, P., Becker, T., Wang, E., ... \& Brody, S. (2025). Improving Large Language Model Performance Through Compression and Optimization. Authorea Preprints.

\item \label{wang2025empowering}
Wang, R., Gao, Z., Zhang, L., Yue, S., \& Gao, Z. (2025). Empowering large language models to edge intelligence: A survey of edge efficient LLMs and techniques. Computer Science Review, 57, 100755.

\item \label{cheng2024small}
Subramanian, S., Elango, V., \& Gungor, M. (2025). Small Language Models (SLMs) Can Still Pack a Punch: A survey. arXiv preprint arXiv:2501.05465.

\item \label{cheng2024survey}
Cheng, H., Zhang, M., \& Shi, J. Q. (2024). A survey on deep neural network pruning: Taxonomy, comparison, analysis, and recommendations. IEEE Transactions on Pattern Analysis and Machine Intelligence.

\item \label{zhu2024survey}
Zhu, X., Li, J., Liu, Y., Ma, C., \& Wang, W. (2024). A survey on model compression for large language models. Transactions of the Association for Computational Linguistics, 12, 1556-1577.

\item \label{tang2024survey}
Tang, Y., Wang, Y., Guo, J., Tu, Z., Han, K., Hu, H., \& Tao, D. (2024). A survey on transformer compression. arXiv preprint arXiv:2402.05964.

\item \label{xu2020bert}
Xu, C., Zhou, W., Ge, T., Wei, F., \& Zhou, M. (2020). Bert-of-theseus: Compressing bert by progressive module replacing. arXiv preprint arXiv:2002.02925.
\item \label{subramanian2025small}
Subramanian, S., Elango, V., \& Gungor, M. (2025). Small Language Models (SLMs) Can Still Pack a Punch: A survey. arXiv preprint arXiv:2501.05465.

\item \label{goddard2024mergekit}
Goddard, C., Siriwardhana, S., Ehghaghi, M., Meyers, L., Karpukhin, V., Benedict, B., ... \& Solawetz, J. (2024, November). Arcee’s mergekit: A toolkit for merging large language models. In Proceedings of the 2024 Conference on Empirical Methods in Natural Language Processing: Industry Track (pp. 477-485).
\item \label{xu2024survey}
Xu, X., Li, M., Tao, C., Shen, T., Cheng, R., Li, J., ... \& Zhou, T. (2024). A survey on knowledge distillation of large language models. arXiv preprint arXiv:2402.13116.

\item \label{zhao2023survey}
Zhao, W. X., Zhou, K., Li, J., Tang, T., Wang, X., Hou, Y., ... \& Wen, J. R. (2023). A survey of large language models. arXiv preprint arXiv:2303.18223, 1(2).

\item \label{hadi2023large}
Hadi, M. U., Qureshi, R., Shah, A., Irfan, M., Zafar, A., Shaikh, M. B., ... \& Mirjalili, S. (2023). Large language models: a comprehensive survey of its applications, challenges, limitations, and future prospects. Authorea Preprints, 1, 1-26.

\item \label{lu2024small}
Lu, Z., Li, X., Cai, D., Yi, R., Liu, F., Zhang, X., ... \& Xu, M. (2024). Small language models: Survey, measurements, and insights. arXiv preprint arXiv:2409.15790.

\item \label{minaee2025largelanguagemodelssurvey}
Minaee, S., Mikolov, T., Nikzad, N., Chenaghlu, M., Socher, R., Amatriain, X., \& Gao, J. Large language models: A survey. arXiv 2024. arXiv preprint arXiv:2402.06196.

\item \label{Dong2024inference}
Dong, X., Teleki, M., \& Caverlee, J. (2024). A Survey on LLM Inference-Time Self-Improvement. arXiv preprint arXiv:2412.14352.

\item \label{biesialska2020continual}
Biesialska, M., Biesialska, K., \& Costa-Jussa, M. R. (2020). Continual lifelong learning in natural language processing: A survey. arXiv preprint arXiv:2012.09823.

\item \label{li2024federated}
Li, L., Gou, J., Yu, B., Du, L., \& Tao, Z. Y. D. (2024). Federated distillation: A survey. arXiv preprint arXiv:2404.08564.

\item \label{shao2023survey}
Shao, J., Li, Z., Sun, W., Zhou, T., Sun, Y., Liu, L., ... \& Zhang, J. (2023). A survey of what to share in federated learning: Perspectives on model utility, privacy leakage, and communication efficiency. arXiv preprint arXiv:2307.10655.

\item \label{albalak2024survey}
Albalak, A., Elazar, Y., Xie, S. M., Longpre, S., Lambert, N., Wang, X., ... \& Wang, W. Y. (2024). A survey on data selection for language models. arXiv preprint arXiv:2402.16827.

\item \label{liu2024datasets}
Liu, Y., Cao, J., Liu, C., Ding, K., \& Jin, L. (2024). Datasets for large language models: A comprehensive survey. arXiv preprint arXiv:2402.18041.

\item \label{ip2025llm}
Ip, J. (2024). LLM Evaluation Metrics: The Ultimate LLM Evaluation Guide. Confident AI.

\item \label{microsoft2024llmeval}
Microsoft Learn. (2024). A list of metrics for evaluating llm-generated content. Microsoft.

\item \label{xu2023parameter}
Xu, L., Xie, H., Qin, S. Z. J., Tao, X., \& Wang, F. L. (2023). Parameter-efficient fine-tuning methods for pretrained language models: A critical review and assessment. arXiv preprint arXiv:2312.12148.

\item \label{wang2019haq}
Wang, K., Liu, Z., Lin, Y., Lin, J., \& Han, S. (2019). Haq: Hardware-aware automated quantization with mixed precision. In Proceedings of the IEEE/CVF conference on computer vision and pattern recognition (pp. 8612-8620).

\item \label{zhang2023loraprune}
Zhang, M., Chen, H., Shen, C., Yang, Z., Ou, L., Yu, X., \& Zhuang, B. (2023). LoRAPrune: Structured pruning meets low-rank parameter-efficient fine-tuning. arXiv preprint arXiv:2305.18403.

\item \label{thomas2024multi}
D. Thomas, D. Maniloff, \& D. Holtz. (2024) Tgi multi-lora: Deploy once, serve 30 models. Hugging Face Blog.

\item \label{miao2023towards}
Miao, X., Oliaro, G., Zhang, Z., Cheng, X., Jin, H., Chen, T., \& Jia, Z. (2023). Towards efficient generative large language model serving: A survey from algorithms to systems. arXiv preprint arXiv:2312.15234.

\item \label{wang2024model}
Wang, W., Chen, W., Luo, Y., Long, Y., Lin, Z., Zhang, L., ... \& He, X. (2024). Model compression and efficient inference for large language models: A survey. arXiv preprint arXiv:2402.09748.

\end{enumerate}
}

\newpage
\section*{Appendix B: Complete Methodology}
\label{sec:AppB}

Initially, this study was designed following the established guidelines for Multivocal Literature Reviews (MLRs) as outlined by Garousi et al.~\cite{MLRguidelines}, with the intent to integrate insights from both peer-reviewed and grey literature in a systematic manner. However, during the early stages of literature gathering, we identified a substantial and rapidly growing body of grey literature relevant to our research objective—defining the lifecycle of Small Language Models (SLMs). This emergent body of work, characterized by tooling documentation, practical frameworks, and implementation guides, demonstrated a level of technical depth and practical relevance that would not be captured through peer-reviewed sources alone.

As a result, we strategically revised our approach to conduct a comprehensive survey, aiming for inclusivity and breadth without being constrained by the formal boundaries of a classic MLR. This pivot enabled us to synthesize a broader landscape of current practices and insights, encompassing both scholarly and practitioner perspectives. To ensure methodological transparency and replicability, we structured the survey process using MLR-inspired rigor, while allowing for the flexible integration of informal sources.

In the following subsections, we present the structure of our approach, including: the research goal and questions (Section~\ref{RQ}), our literature search strategy (Section~\ref{Strategy}), the quality assessment process (Section~\ref{QualityAssessment}), and our data extraction and analysis procedures (Section~\ref{DataExtraction}). Finally, we reflect on the replicability of our review given the inclusion of non-traditional sources (Section~\ref{Replicability}).

\subsection{Goal and Research Questions}
\label{RQ}

The overarching goal of this study is to define a comprehensive lifecycle for Small Language Models (SLMs) that integrates techniques and practices from both academic and industrial contexts. Our investigation is guided by two research questions:

\begin{itemize}
    \item \textbf{RQ1:} How can a comprehensive, end-to-end lifecycle for Small Language Models be defined?
    \item \textbf{RQ2:} What are the key components and interfaces of the SLM lifecycle, and how do they interact?
\end{itemize}

RQ1 focuses on lifecycle structuring, aiming to synthesize diverse contributions into a cohesive framework. RQ2 addresses the dynamic interconnections between components, with particular attention to modularity, interoperability, and method reuse.

\subsection{Search Strategy}
\label{Strategy}

Our search strategy combined structured querying with iterative snowballing and expert-informed discovery. We selected a core set of survey papers published between 2022 and 2025 that explicitly focus on SLMs, LLMs, and efficiency techniques in language models. Bibliographic databases queried include Google Scholar, arXiv, and Scopus using combinations of the terms:

\textit{"Small Language Models", "SLM lifecycle", "efficient fine-tuning", "quantization LLM", "distillation LLM", "LLM deployment", "SLMOps"}.

To enrich our understanding of emerging practices, we also included technically detailed blog posts, library documentation, and community reports retrieved from Medium, GitHub repositories, mailing lists, newsletters, and curated repositories. In total, we selected a total of 36 sources.

\subsection{Quality Assessment}
\label{QualityAssessment}

Given the methodological diversity of our sources, we applied a lightweight quality screening process adapted from MLR guidelines. For academic papers, inclusion was based on relevance to SLM lifecycle stages, clarity of scope, and citation impact. For grey literature, we prioritized sources that offered implementation-level detail, referenced empirical observations, or were authored by domain experts or core contributors to key open-source tools. All selected sources were reviewed by at least two authors to ensure relevance and content quality.

\subsection{Data Extraction and Analysis}
\label{DataExtraction}

Data extraction was guided by our two research questions. For RQ1, we classified lifecycle-relevant practices and concepts into three categories: \textbf{main components}, \textbf{optional components}, and \textbf{cross-cutting elements}. For RQ2, we identified interconnections between components, including method reuse, interface dependencies, and co-adaptation strategies (e.g., quantization-aware fine-tuning, PEFT-aware pruning). Data synthesis was carried out through iterative coding, informed by lifecycle modeling and thematic clustering.

The final lifecycle framework was validated through author consensus and cross-checked against recent tooling documentation to ensure technical soundness.

\subsection{Replicability}
\label{Replicability}

While this study maintains a structured and transparent methodology, the inclusion of evolving, grey sources limits full reproducibility. Initially, our source retrieval followed established guidelines for MLRs, using defined search strings and selection protocols. However, as the study progressed, we observed that many of the most relevant and practically informative works emerged through less formal channels, such as community-curated repositories, newsletters, Medium posts, and tooling documentation.

Given the dynamic and practitioner-led nature of the field, we made a conscious decision to abandon a strict MLR protocol and adopt a comprehensive survey approach. This shift allowed us to prioritize conceptual depth and relevance over formal search traceability. As a result, we opted not to construct a formal replication package, as it would not meaningfully reflect the emergent, iterative way in which key sources were identified and incorporated.

Instead, to maintain transparency, we provide a categorized list of all sources used in this study (including informal materials) in Appendix~A, along with annotations that clarify their relevance and role in shaping the proposed lifecycle framework.
\end{document}